# Evidence of near-infrared partial photonic bandgap in polymeric rod-connected diamond structures

Lifeng Chen[1], Mike P. C. Taverne[1], Xu Zheng[1], Jia-De Lin[1], Ruth Oulton[1,2], Martin Lopez-Garcia[1], Ying-Lung D. Ho,[1,3] and John G. Rarity[1,4]

[1]*Department of Electrical and Electronic Engineering, University of Bristol, Merchant Venturers Building, Woodland Road, Bristol BS8 1UB, UK*
[2]*H.H. Wills Physics Laboratory, University of Bristol, Tyndall Avenue, Bristol BS8 1FD, UK*
[3]*Daniel.Ho@bristol.ac.uk*
[4]*John.Rarity@bristol.ac.uk*

**Abstract:** We present the simulation, fabrication, and optical characterization of low-index polymeric rod-connected diamond (RCD) structures. Such complex three-dimensional photonic crystal structures are created via direct laser writing by two-photon polymerization. To our knowledge, this is the first measurement at near-infrared wavelengths, showing partial photonic bandgaps for this structure. We characterize structures in transmission and reflection using angular resolved Fourier image spectroscopy to visualize the band structure. Comparison of the numerical simulations of such structures with the experimentally measured data show good agreement for both P- and S-polarizations.

## 1. Introduction

Three-dimensional (3D) photonic crystals (PhCs) with complete or partial photonic bandgaps (PBGs) have attracted keen interest from academia and industry since 1987 [1]. In recent years, much effort has been devoted to the fabrication of 3D PhCs in the near-infrared communication band (1.3-1.7 µm) [2–9] and the visible wavelength range [10]. Among these, fabrication of 3D structures by direct laser writing (DLW) using two-photon polymerization (TPP) [11] has been studied extensively due to its great flexibility and its capability to precisely recreate complex geometries. The DLW technique allows fabrication of any arbitrarily designed 3D structure in a photosensitive polymeric material by scanning a tightly focused pulsed laser spot across the photoresist. The TPP method enhances the resolution providing lateral features down to 100 nm [12–14]. This has led to fabrication and demonstration of woodpile PhCs [15] with extensions to controlled spontaneous emission [16], invisibility cloaking [17], and chirality [18,19].

Recently, light confinement in photonic diamond lattice structures such as rod-connected diamond (RCD) [20-23], have been investigated via the plane-wave expansion (PWE) [24] method and finite-difference time domain (FDTD) calculations [25]. The main advantages of the the RCD structure over more conventional woodpile designs are (a) a lower minimum refractive index contrast required for a full PBG of less than 1.9 [20] (>1.9 for fcc woodpile structures [26]), (b) the largest gap width to centre frequency (wavelength) ratio $\Delta\omega/\omega_0$ reaching >30% for index contrast 3.6 [23] (woodpile is ~18%), and (c) higher mid-gap centre frequency making fabrication on the scale of optical wavelengths easier. This makes it more fault-tolerant and thus in principle easier to fabricate, despite the more complex geometry.

Woodpile structures can be fabricated using layer by layer lithographic approaches [27–29] and infrared bandgaps have been measured [29]. This approach is not suited to RCD structures and fabrication of such intricate 3D structures at arbitrary length scales, particularly at near infrared communications wavelengths (1.4-1.6 µm), still remains a significant challenge. Naturally occurring biological PhC structures similar to RCD (e.g. *Entimus Imperialis*, diamond weevil) have been used as templates for visible bandgap materials [30,31] but at wavelengths fixed by the template.

In our work, we are developing methods to manufacture arbitrary scale photonic bandgap materials using polymer templates using 3D TPP lithography. Here we report on RCD structures directly laser-written in polymer ($n = 1.52$) that show partial photonic bandgaps and evidence of 3D photonic bandstructure. A range of advanced optical modelling tools are used to provide data to guide structure design, which allow us to predict and interpret their performance post-fabrication. Angular resolved Fourier imaging spectroscopy measurements of transmission and reflection are used to characterize the bandstructure of the fabricated structures [32,33]. These results are compared with simulations and discrepancies between the experimental data and simulations used to improve fabrication to the point where working 3D RCD structures could be demonstrated.

## 2. Experiments and description

The RCD structure replaces the bonds in a diamond lattice with rods as shown in Fig. 1(a). This can be visualized as a non-primitive cubic unit cell in Fig. 1(b). It contains 16 elliptical cylinders which can be grouped into four regular tetrahedral arrangements. The translational symmetries are the same as for a face-centered cubic (FCC) structure, with a primitive unit-cell only containing one of these tetrahedral arrangements. The corresponding reciprocal lattice is a body-centered cubic (BCC) lattice, leading to a first Brillouin zone which is a truncated octahedron with BCC symmetry as illustrated in red in Fig. 1(b).

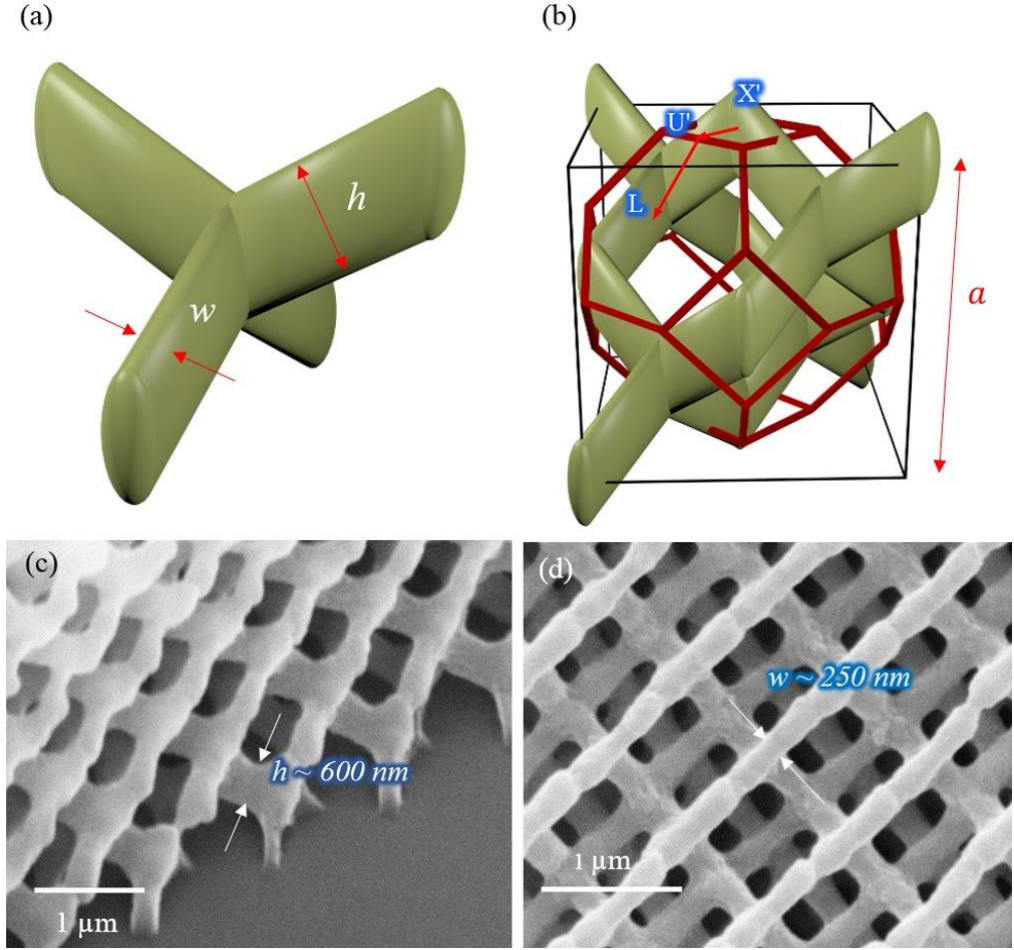

Fig. 1. (a) Tetrahedral unit cell showing width ($w$) and height ($h$) of rod. (b) Brillouin zone of RCD, $a$ is the lattice constant for a unit cell. SEM images of fabricated structures (c) and (d) show the feature sizes: (c) rod height $h \sim 600$ nm and (d) rod width $w \sim 250$ nm.

The DLW system we used in fabrication is a commercial instrument based on the TPP technique (Photonic Professional, Nanoscribe GmbH), which contains a 780 nm femtosecond laser beam (with pulse width ~ 120 fs and repetition rate ~ 80 MHz) focused into a negative photoresist (IP-L 780, Nanoscribe GmbH) by a high numerical aperture (NA = 1.4) oil immersion objective lens (100×, Zeiss). Such liquid photoresist formulation (with refractive index $n = 1.52$ after polymerization) provides a high resolution (single voxel size) down to 150 nm in the horizontal plane and vertical: horizontal aspect ratio ~ 3:1. We chose the diamond lattice constant (the length of a side of the cubic unit-cell) $a = 1.25$ μm so that the fundamental partial band gap in our measureable incident angle range appears for a wavelength between 1.4 μm and 1.6 μm. The voxel shape in the fabrication process is elliptical and we label the width and height of rods as *w* [Fig. 1(a)]. With multiple exposure on a regular scanning path , we achieve $h$ = 600 nm and $w$ = 250 nm as seen in the scanning electron microscope (SEM) images shown in Figs. 1(c) and 2(d). This results in a filling fraction of approximately 48%. Each structure fabricated is free standing on a standard coverslip. The structure size, we made for test, is approximately 14 μm × 14 μm in plane and 7.5 μm vertically [Fig. 2(a)]. The laser profile at the focal point varies as focus depth changes due to a refractive index mismatch between the photoresist and the glass substrate [5]. Additionally, shrinkage occurs during polymerization due to the nature of the polymer material itself [34]. These phenomena make the aspect ratio of voxels changes as the structure grows higher, with approximately 25nm decrease in rod height per layer (approx. 3.5% per layer). As a result of this non-uniform feature size a chirped RCD structure is formed, as shown in Figs. 2(b) and 2(c), which influences the photonic band diagram. The consequences of this imperfection will be discussed later.

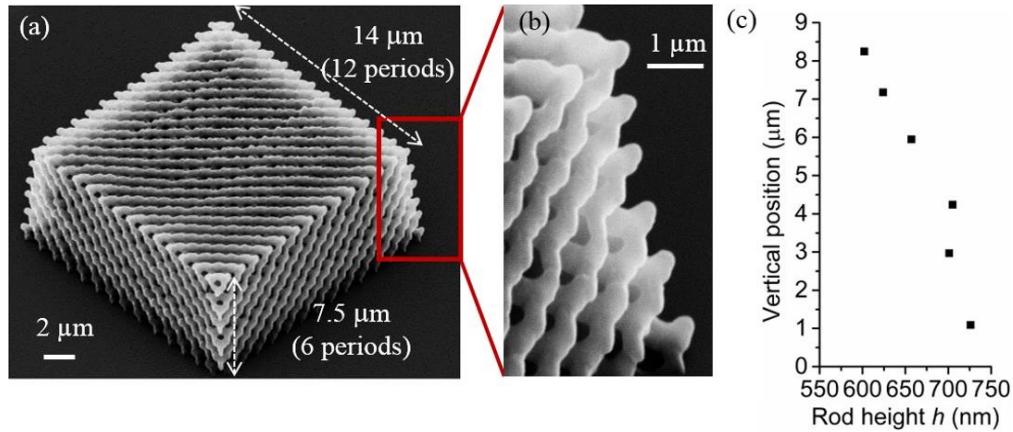

Fig. 2. (a) SEM image of the overview of a whole RCD structure: 12 periods in plane (14 μm) and 6 periods in vertical direction (7.5 μm). Detailed side view image zoomed in (b) shows the non-uniform feature sizes formed with a chirped RCD structure. (c) Illustrates the rod height (*h*) against the vertical position of this 6 layers RCD structure. At the top layer $h \sim$ 600 nm while the bottom layer $h \sim$ 725 nm suggesting $\sim$ 25 nm increment of rod height, per layer.

The measurement setup is based on the Fourier imaging spectroscopy principle [35] where reflection spectra are collected at various angles in the Fourier plane of a lens. Figure 3 shows a sketch of the setup. A broad band white light source (WLS100 from Bentham Instruments Ltd.) is coupled to a multimode fiber to flexibly transport light to the measurement system. The output light is collimated by a low magnification objective lens (4×), passed through a linear polarizer and focused by a high numerical aperture (40× NA = 0.75) objective lens onto the sample. Reflected light is collected by the same objective lens and picked off in a 50% beam splitter followed by a 4-f system (L2 and L3), focused on the back focal plane of the objective lens. This then transfers the Fourier image onto the detection plane  where light is collected into a fiber mounted on a motorized scanning stage (with spatial resolution 1μm per step). The detecting fiber is connected to a spectrometer with an InGaAs photodiode array detector sensitive to a wavelength range from 900 nm to 1800 nm with spectrum resolution 1.5 nm. By scanning the fiber across the Fourier image in the X direction (parallel to the set-up plane), one can construct a spectral map with intensity information for each wavelength collected for every angle in the X plane. The angular resolution partly depends on the detector fiber (normally a 200 micron diameter collects ~3° at each position); the objective lens we use (NA = 0.75) provides a maximum 48° collection angle in theory. For the polarization measurements, a linear polarizer with working wavelength from 750 nm to 2000 nm is placed in front of the beam splitter to introduce light polarized perpendicular to the plane of incidence (S-polarized light) or light polarized parallel to the plane of incidence (P-polarized light).

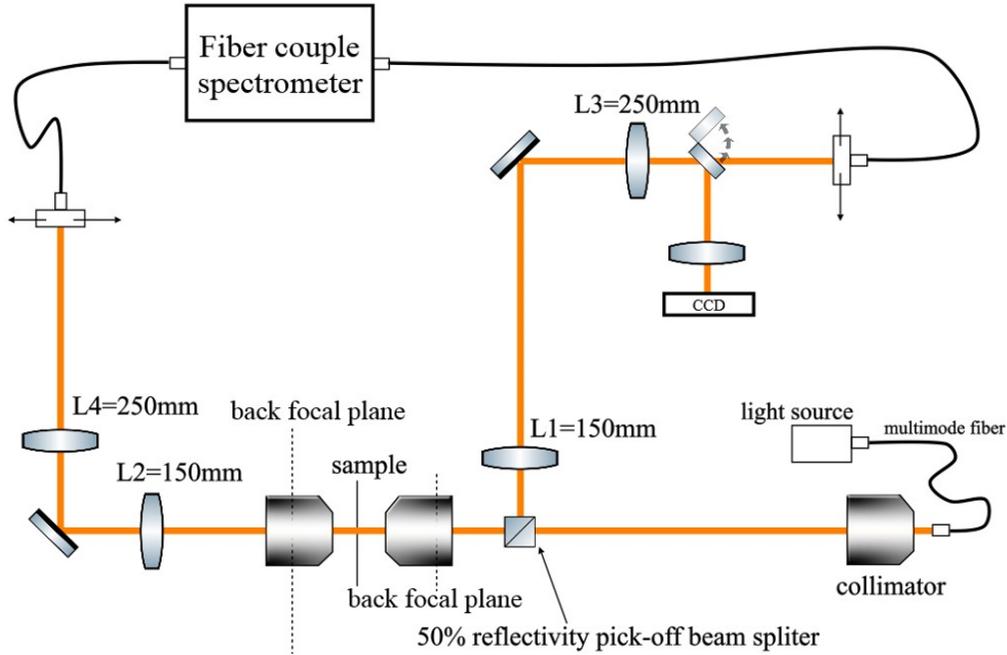

Fig. 3. Diagram of reflection spectroscopy setup, CCD camera is added for orientation on the sample substrate. Lenses L1 and L3 (L2 and L4) relay the back focal plane to the detection fiber in reflection (transmission).

## 3. Results and discussions

To match our expectation and the measurement of fabricated RCD structures, the photonic band diagram of RCD is calculated by the MIT Photonic Bands (MPB) electromagnetic wave solver [24] using PWE method using parameters from SEM measurements [Figs. 1(c) and 1(d)]. In Fig. 4(a), the insert shows the bandstructure calculation within the Brillouin zone, with the blue arrow indicating the scanning direction in measurement. The band diagram shown in Fig. 4(a) has been converted from K-space to angle, i.e. with 0°, 19° and 60° corresponding to Γ-X', Γ-U', and Γ-L. Transmission and reflection spectra are normalized to unity by dividing the data by equivalent spectra taken from glass substrate (transmission) and gold mirror (reflection). Figure 4(b) shows transmission and reflectivity measurements at normal incidence (the X' symmetry point at 0°). A clear 20% dip (peak) in transmission (reflection) with center wavelength around 1500 nm is observed in these measurements. The agreement between simulation and experiment confirms the evidence of the fundamental PBG in this low-index-contrast (1.52 : 1) polymeric RCD structure.

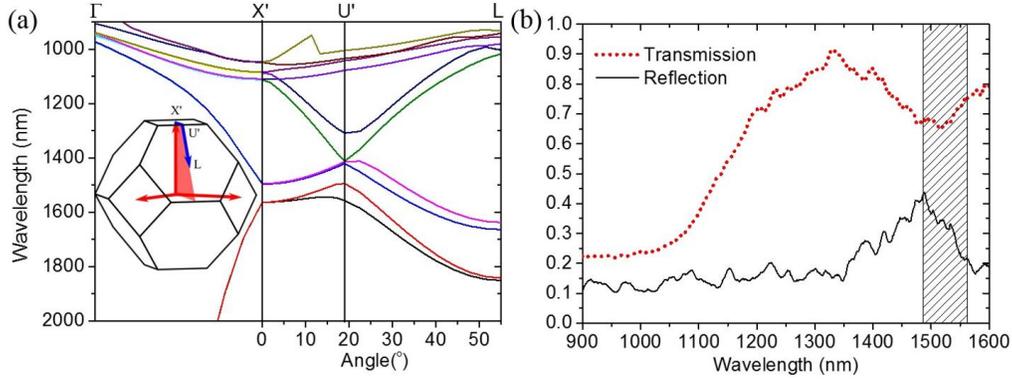

Fig. 4. (a) MIT photonic-bands (MPB) calculation of RCD bandstructure (wavelength against wavevector) showing our measurement range in terms of scattering angle. The insert defines the Brillouin zone of RCD showing the principle directions (X', U', L) with the path of wavevectors covered in our measurements X'→U'→L mapped by bold blue vectors. (b) Transmission and reflection measurements at normal incidence [the X' symmetry point at 0° in Fig. 3(a)] as a function of wavelength. The shadow region defines the relevant fundamental band gap estimated from Fig. 3. (a).

Figures 5(a) and 5(c) show angle resolved optical spectra in reflection for P- (parallel to the incident plane) and S- (perpendicular to the incident plane) polarization with superimposed band diagrams [from Fig. 4(a)]. Figures 5(b) and 5(d) show the FDTD simulation of a finite thickness (7µm), infinite width, chirped (explained later) RCD model that results in the reflectivity stop band ~ 30%, in normal incidence. Both numerical simulation and optical measurement results demonstrate a fundamental partial PBG with center wavelength around 1.55 µm for normal incidence. Moreover, the fundamental PBG moves to higher frequencies as the incident angle increases up to around 20°. Thus, for the case of S-polarization, a strong reflection peak at around 20° is captured in measurement results [Fig. 5(c)], which indicates the turning point U' in the Brillouin zone as shown in Fig. 4(a). For the larger angles (>20°), the fundamental PBG shifts to lower frequencies, but it becomes difficult to observe in the experiment above 30° due to the finite-size [Fig. 2(a)] the crystal in the horizontal plane. Light strongly scattered from the sample edge dominates at large incident angle (>30°). Furthermore, it is difficult to observe the higher-order bands. Nevertheless, a thin faint line extending from the top right corner (near 35° at around 1000 nm) to 0° at around 1100 nm for P-polarized light [Fig. 5(a)] is apparent. Similarly, for S-polarized light [Fig. 5(c)], a line appears near 20° at 1400 nm and extends to the upper left near 0° at 1100 nm. These measurements agree well with the simulations [Figs 5(b) and 5(d)]. The generally poor signal-to-background ratio in the measured spectra comes mainly from light scattering from sample imperfections such as sub-wavelength surface roughness.

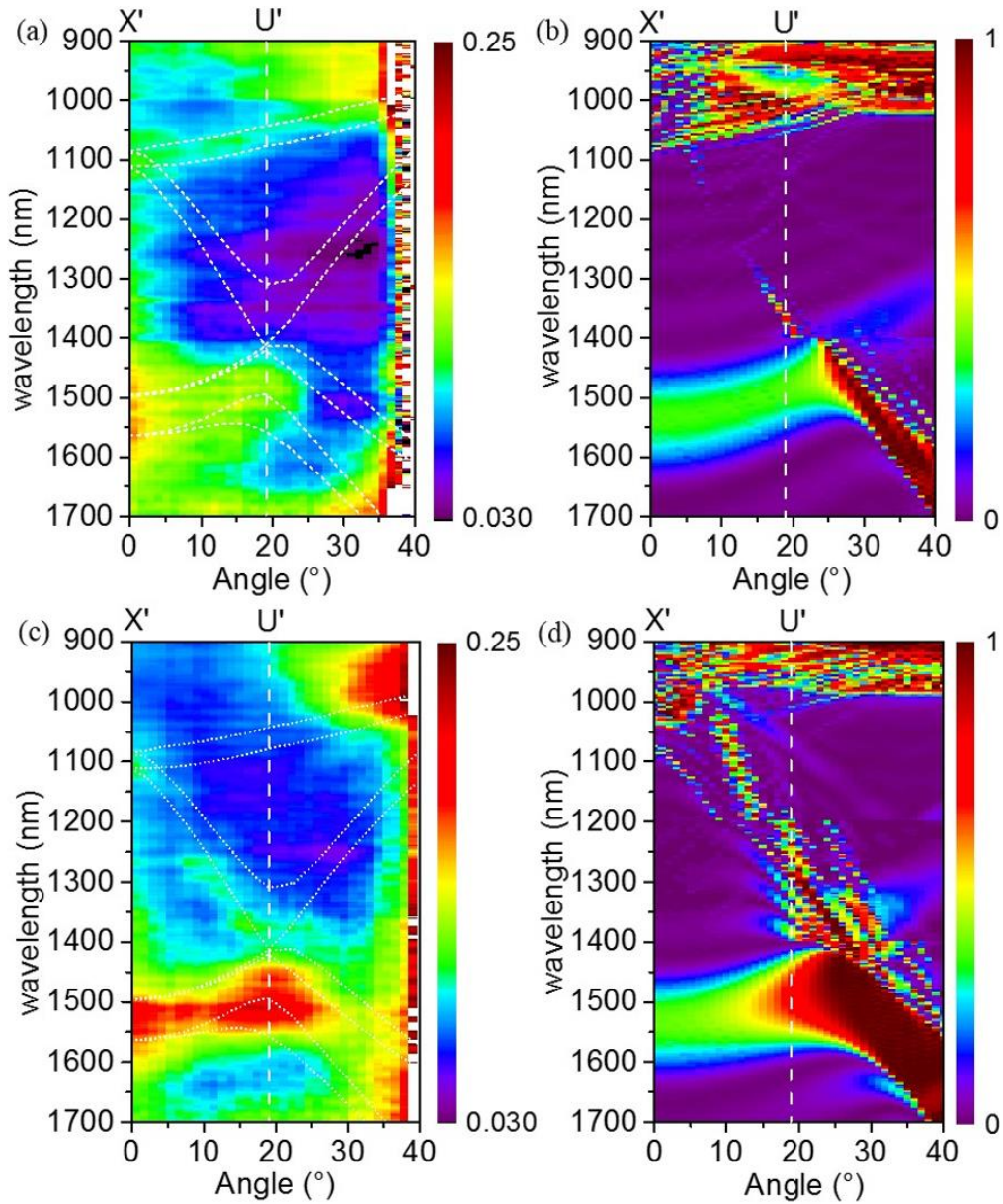

Fig. 5. False color plots of reflection spectra against collection angle (red is high intensity, blue is low) allowing us to visualize bandstructure. (a) Measured and (b) calculated using P-polarized incident light. (c) Measured and (d) calculated using S-polarized light.

As described earlier, the RCD structures fabricated by DLW are imperfect. In Fig. 2(b), a chirped RCD structure is seen, leading to a gradual change of the filling fraction and a shift of bandgap. To illustrate the effect of this on our results, we present here a series of numerical calculations. Figure 6 illustrates how rod size variation influences the bandgap at normal incidence using the FDTD simulations, where the fundamental dimensions of the structure are $h = 600$ nm and $w = 250$ nm with 6 periods in the vertical direction. In Fig. 6(b), the fundamental PBG peak is red-shifted by ~20 nm for every 10 nm increment of rod width ('$\delta w$'). Figure 6(a) shows the fundamental PBG peak is red-shifted by ~10 nm for every 10 nm increment of rod height ('$\delta h$'), but more importantly the reflectivity drops significantly from 50% to 25%.

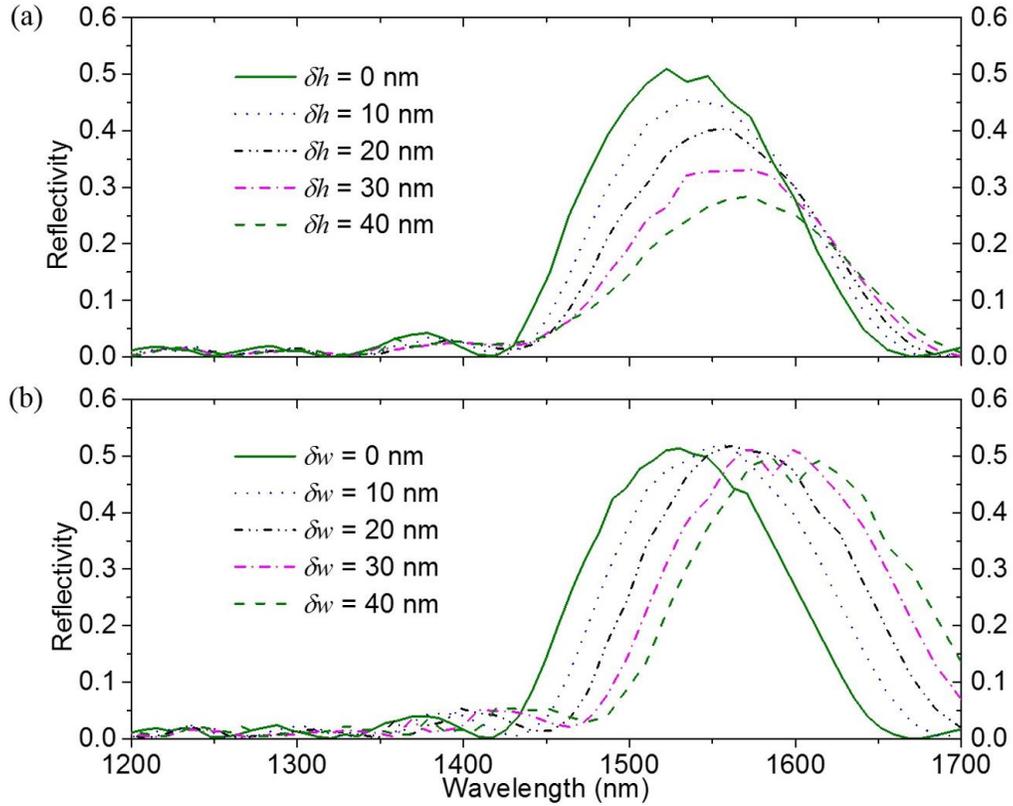

Fig. 6. FDTD simulation of normal incidence reflectivity as a function of wavelength for a chirped RCD structure. (a) Rod height incremented by $\delta h$; (b) rod width incremented by $\delta w$ for each additional layer.

Figure 7 shows a comparison between the FDTD simulation and optical measurement results for 0° (normal), 10° and 19° (Γ-U' direction) incidence angle, where the data is sampled from Fig. 5. We find that there is a good match of the center frequency of the reflection peak for all incidence angles. However, a low reflectivity (18% and 22%) is observed for the measurement for all angles of incidence and a higher background light level (10%) is seen at higher frequencies (i.e. short wavelengths). These issues are partly a result of the chirped RCD structure, fabrication errors, and surface roughness. The simulation at 19° shows strong oscillations at higher frequencies (1200 to 1400 nm wavelength) due to resonant coupling to guided modes (surface or bulk). These are not seen in measurements due to the imperfections in fabrication, but may be a source of the high background.

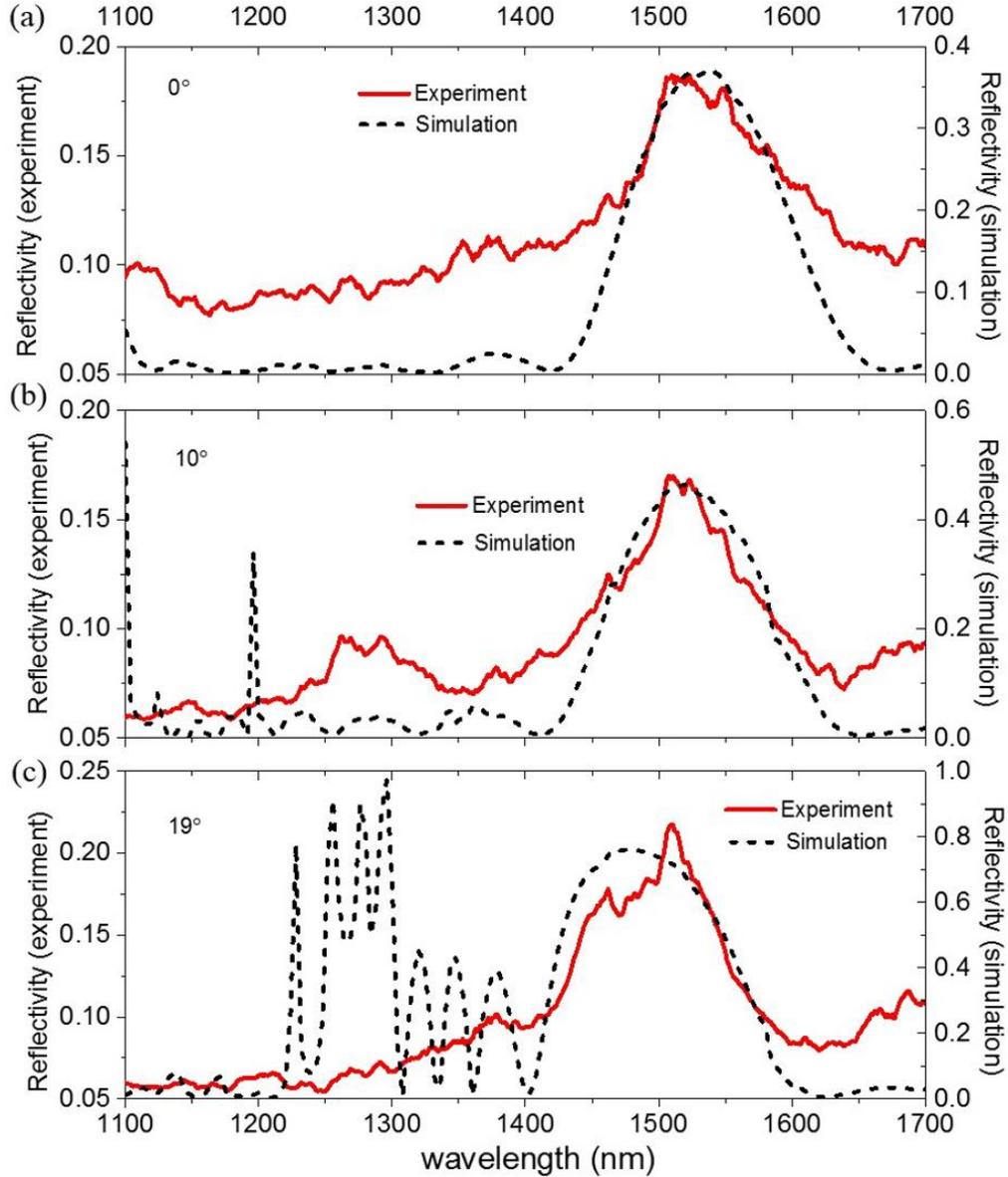

Fig. 7. Plots of reflectivity against wavelength comparing simulations (dashed lines) and experiments (bold lines) at incident angle of 0°, 10° and 19° in (a), (b) and (c).

## 4. Conclusions

In conclusion, we have numerically and experimentally shown evidence of the partial fundamental PBG in a low-index-contrast polymeric RCD structure. We have good control of the size of rod, width ($w$ = 250 nm) and height ($h$ = 600 nm). This leads to a filling fraction of a RCD unit cell of ~ 48% and a reflectivity stop band of ~40% at normal incidence for the design wavelength. We have also investigated the non-uniform feature size which leads to a chirped RCD structure from the TPP process. This chirp leads to a red-shift of the fundamental PBG peak and reduced reflectivity. Despite the chirping phenomenon, there is a good match between the wavelength reflection peak and angle of incidence at 0° and 19°, where we measure reflectivities of 18% and 22% respectively.

For most 3D PhCs suitable for the commercially relevant 1.55 μm wavelength, the desired voxel aspect ratio is 1.

In future work, we will investigate various ways to reduce this ratio, including the use of a spatial light modulator (SLM) for adaptive beam shaping and dark field illumination schemes [36,37]. The stimulated emission depletion (STED) technique [38,39] could also be configured to improve aspect ratio although is best for improving lateral resolution. Switching to shorter write wavelengths [40] will reduce the minimum feature size including voxel depth thus allowing more flexibility of near infra-red structure design [19].

To the best of our knowledge, we believe this to be the first example of a 3D DLW fabricated polymeric RCD photonic crystal showing partial PBGs at near-infrared wavelength (1.55 μm). A future aim is to engineer high-refractive-index-contrast wavelength scale structures with complete PBGs. This can be done either by backfilling structures fabricated in photoresist using high-index material or by DLW in high refractive index photosensitive materials such as chalcogenides [41]. In the longer term, we aim to develop defect cavities [42, 43] and investigate cavity QED effects [44, 45] with single emitters such as quantum dots [46] or color centers [47].

**Acknowledgments**

This work was carried out using the computational facilities of the Advanced Computing Research Centre and the nanofabrication and characterization equipment of the Centre for Nanoscience and Quantum Information, University of Bristol, Bristol, U. K. We acknowledge financial supports from the ERC advanced grant 247462 QUOWSS and the EU FP7 grants 618078 WASPS and 284743 SPANGL4Q and the EPSRC grant EP/M009033/1.